\documentclass[preprint, onecolumn,12pt]{elsarticle}

\usepackage[english]{babel} 
\usepackage{lineno,hyperref}
\usepackage[final]{changes}
\usepackage{ulem}
\usepackage{upgreek}
\usepackage{color}

\journal{Journal of \LaTeX\ Templates}

\graphicspath{{Figures/}{Figures/DesignDoc/}}

\def \CSACO{CSA\textsubscript{w/CO} }
\def \micro#1{\upmu#1}       
\def \Ohm{\Omega}
\def \outQ{\textit{out\_charge} }
\def \outI{\textit{out\_current} }
\def \charge{\textit{charge}}
\def \current{\textit{current}}
\def \nbd{\babelhyphen{nobreak}}

\usepackage{enumitem}
\setlist[itemize]{leftmargin=*}      

\usepackage{tabularx}

\usepackage{todonotes}               

\begin{document}

\begin{frontmatter}

\title{Pulse shape discrimination for GRIT:  beam test of a new integrated charge and current preamplifier coupled with high granularity Silicon detectors.}


\author{}
\address{}

\author[ijc]{J.-J. Dormard}
\author[ijc]{M. Assi\'e}
\cortext[mycorrespondingauthor]{Corresponding author: marlene.assie@ijclab.in2p3.fr}
\author[ijc]{L. Grassi}
\author[ijc]{E. Rauly}
\author[ijc]{D. Beaumel}
\author[ijc]{G. Brulin}
\author[ijc]{M. Chabot}
\author[ijc]{J.-L. Coacolo}
\author[ijc]{F. Flavigny} \fntext[fn1]{\emph{Present address:} Normandie Univ, ENSICAEN, UNICAEN, CNRS/IN2P3, LPC Caen, 14000 Caen, France}
\author[ijc]{B. Genolini}
\author[ijc]{F. Hammache}
\author[ijc]{T. Id Barkach}
\author[ijc]{E. Rindel}
\author[ijc]{Ph. Rosier}
\author[ijc]{N. de S\'er\'eville}
\author[ijc]{E. Wanlin}


\address[ijc]{Universit\'e Paris-Saclay, CNRS/IN2P3, IJCLab, 91405 Orsay, France}

\begin{abstract}

The GRIT (Granularity, Resolution, Identification, Transparency) Silicon array is intended to measure direct reactions. Its design is based on several layers (three layers in the forward direction, two backward) of custom-made trapezoidal and square detectors. The first stage is 500~$\micro$m thick and features 128$\times$128 orthogonal strips. Pulse shape analysis for particle identification is implemented for this first layer. Given the compacity of this array and the large number of channels involved ($>$7,500), an integrated preamplifier, iPACI, that gives \charge{} and \current{} information has been developed in the AMS 0.35~$\micro$m BiCMOS technology. The design specifications and results of the test bench are presented.
\added{Considering an energy range of 50~MeV and an energy resolution (FWHM) of 12~keV (FWHM) for the preamplifier}, the energy resolution for one strip obtained from alpha source measurement in real conditions is 35~keV. The current output bandwidth is measured at 130 MHz for small signals and the power consumption reaches 40~mW per detector channel. 
 A first beam test was performed coupling a nTD trapezoidal double-sided stripped Silicon detector of GRIT with the iPACI preamplifier and a 64-channel digitizer. Z=1 particles are discriminated with pulse shape analysis technique down to 2~MeV for protons, 2.5~MeV for deuterons and 3~MeV for tritons. The effect of the strip length due to the trapezoidal shape of the detector is investigated on both the N- and the P-side, showing no significant impact.
\end{abstract}


\end{frontmatter}


\section{Introduction}

Direct reactions are one of the cornerstones to study nuclear structure and nuclear astrophysics as they provide precise information on the single particle and collective properties of nuclear states. 
The limitation to stable beams in direct kinematics was overcome by the inverse kinematics where the probe is no longer the beam but the target. A wide range of radioactive ion beams becomes available and opens up a complete new era of direct reactions studies.
However inverse kinematics comes with a number of experimental challenges. Some of them have been solved with the Silicon detector technology that enables to identify and measure the energy loss and angle of charged particles with a large angular coverage.
Moreover, the new generation of facilities in Europe, like SPES \cite{spes}, HIE-ISOLDE \cite{isolde} or FAIR \cite{fair} will provide heavy radioactive beams that will raise new experimental challenges. Indeed the nuclei investigated through direct reactions will have a high density of states. The measurement of the light ejectiles only will not be sufficient to identify the populated states and coupling with gamma-rays detectors will be crucial. \added{This is the reason for which} the new generation of Silicon arrays like GODDESS \cite{goddess}, T-REX \cite{trex} and GRIT \cite{grit} are designed to be coupled with state-of-the-art gamma arrays. 

\begin{figure}[h]
	\begin{center}
		\includegraphics[width=\linewidth]{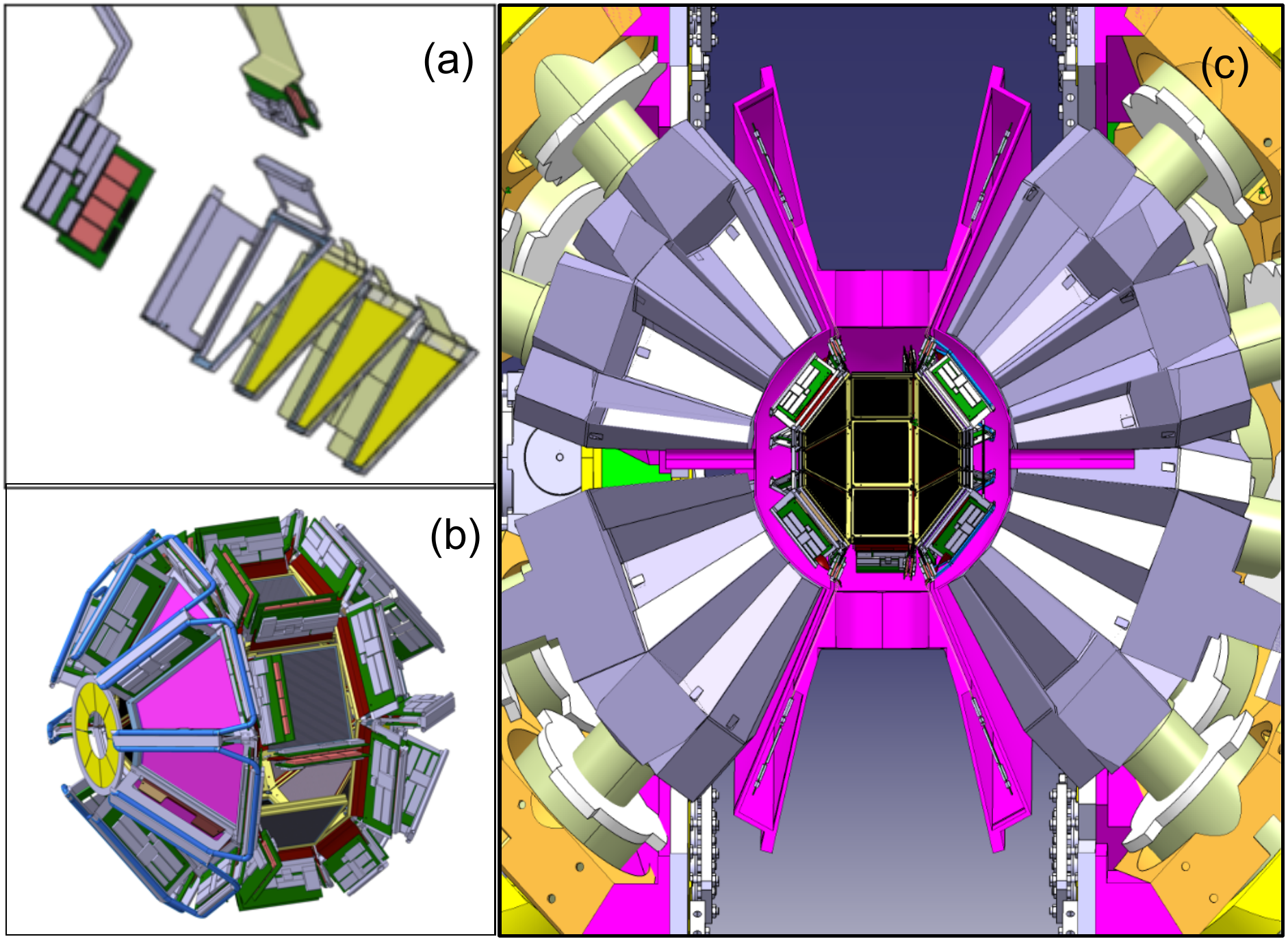} 
		\caption{Preliminary design of the GRIT array: (a) the three Silicon layers and front-end electronics (b) Design of the full array with electronics; (c) GRIT within the AGATA array. }
		\label{design}
	\end{center}
\end{figure}

In this paper we will focus on the developments for the GRIT Silicon array to be coupled with AGATA \cite{agataG}. Its design (see Fig. \ref{design}) fits into a 23-cm radius sphere (the one of AGATA) and is as transparent as possible to gamma-rays. It is composed of two rings (one forward and one backward with respect to the beam direction) of 8 trapezoidal telescopes, each composed of 2 (in the backward hemisphere) to 3 layers (in the forward hemisphere) of double-sided stripped Silicon detectors (DSSD). The detectors feature different thicknesses: 500~$\micro$m for the front layer and 1.5~mm for the others. A ring of 6 square detectors completes the set-up at 90 degrees. With 128 strips on each side for the first layer and 16 for the second and third layers, it amounts to more than 7,500 strips to be read in vacuum with front-end boards attached perpendicularly to the detectors. Table \ref{tab_det} summarizes the detector design of GRIT. The integration of the high-granularity detectors and electronics in a compact reaction chamber under vacuum is a major challenge for GRIT.

\begin{table}
\begin{center}
\begin{tabular}{|c|c|c|c|c|} 
\hline
 & Nb. Tel.  & Thickness & Strips\\ 
\hline \hline
Forward & 8  & 500 $\micro$m & 128X+ 128Y\\ 
              &     & 1500 $\micro$m & 16X+ 16Y\\  
              &     & 1500 $\micro$m & 16X+ 16Y\\ 
              \hline
Backward & 8 & 500 $\micro$m & 128X+ 128Y\\ 
              &     & 1500 $\micro$m & 16X+ 16Y\\ 
              \hline
90 degrees & 6  & 500 $\micro$m & 128X+ 128Y\\ 
              &      & 1500 $\micro$m & 16X+ 16Y\\ 
              \hline \hline
 \end{tabular}
\caption{GRIT detectors description with the number of telescopes per ring and their thicknesses and number of strips.}
\label{tab_det}
\end{center}
\end{table}

In order to overcome the difficulties in particle identification (PId) inherent to the  time-of-flight method used for particles which stop in the first layer of Silicon, the pulse shape analysis (PSA) technique is implemented \cite{Amm63,Eng89,Pau94,Mut00,Mut09,Cha02}. It relies on the shape of the current and/or charge signal to identify the particles. For a given energy, the lighter the particle the faster the rise time and the smaller the width of the current signal. This feature is very useful when the flight path of the particle is too short or when the instantaneous beam intensity is too high to have a time reference of good resolution ($<$ 1.5~ns) like for beam tracking devices.
Following the studies from the FAZIA collaboration \cite{Bard11,Car12, Bar13,LeN13,Val19}  \added{performed for pulse shape discrimination of heavy nuclei on a wide dynamic range (up to GeV), several studies of the PSA focussing on light particles} have been carried out for GRIT \cite{psa,he3,Due12,Gen13,Men14}. They have shown that the optimal observable for the PId is the maximum of the \current{} signal. Other observables can be used like the rise time of the \charge{} signal at the expense of the particle discrimination quality \cite{psa}. 


\added{Considering the large number of channels of the GRIT apparatus and the necessity to have both charge and current information, a new integrated preamplifier has been developed, the so-called iPACI (integrated Pre-Amplifier for Charge and current). Its first version featuring 9 channels and an energy range of 50 MeV is presented in Section \ref{sec_ipaci}. Its specification is described in Section \ref{sec_design}, its architecture  in Section \ref{sec_archi}, while the simulated performance is detailed in Section \ref{sec_simu} and the test results using a pulser in Section \ref{sec_test}. In Section \ref{sec_beam}, the results of a follow-up test experiment where this new circuit is coupled with a trapezoidal DSSD prototype of the GRIT project is presented. Note that previous tests had been performed with a neutron transmutation doped (nTD) 500 $\micro$m thick square DSSD (62x62 mm$^2$, 128x128 strips) designed specifically for the R\&D studied of GRIT using an older version of the preamplifier, the discrete PACI preamplifiers \cite{Ham04}. So, this paper represents a very important additional step because it shows the results of the first beam test using a prototype of the first-layer trapezoidal shaped detector of GRIT read out by the first version of the integrated preamplifier iPACI. In Section \ref{sec_beam}, the Pulse Shape Analysis (PSA) results and the effect of the strip length on the particle discrimination will be presented.}


\section{iPACI design}
\label{sec_ipaci}

The main integration constraint of the GRIT project is to implement 15,000 electronic channels (\charge{} and \current{} signals, related to 7,500 detector strips), in a very compact reaction chamber. 

Designing an integrated preamplifier featuring a \current{} output aimed at PSA \added{purpose} poses some challenges. The first of them is that the system bandwidth has to be high enough, while the noise kept low so that the detector current is accurately reproduced at the \current{} output. Additionally, feeding fast-edged pulses to the outside of chips can lead to cross-talk and instabilities, due to the high current peaks flowing through the supplies.

The work presented below is the first prototype of the iPACI chip where the same architecture as the long-proven PACI design \cite{Ham04} was kept. Indeed, the usual three-stage architecture of discrete Charge-Sensitive Amplifier with Current Output circuits (abbreviated \CSACO below) was used, with independent \charge{} and \current{}, buffered outputs. Keeping the minimum chip size to be sent to fabrication, 4~mm$^2$, the chip can accommodate nine channels. 

\subsection{Specifications}
\label{sec_design}

The first version of the iPACI chip is designed to read \added{the first DSSD layer of the GRIT apparatus,} where PSA will be performed. GRIT is mostly intended to detect light particles (Z$\leq$2) and the $^4$He particles punch through 500~$\micro$m Silicon at 33~MeV. The energy range for the first version of iPACI is 0.2 to 50~MeV, which corresponds to 8.9 to 2200~fC. The energy resolution is required to be of the order of 30~keV (FWHM) per strip, which is a typical value for DSSD. This implies below 10~keV resolution for the \charge{} output after ideal shaper and at optimal time constant.
The detection of light nuclei up to $^{12}$C, which punch through at an energy of 182~MeV, is also of interest. This is why two switchable ranges will be implemented for the final version: 0.2 to 50~MeV and 0.2 to 200~MeV. \added{Increasing the dynamic range while keeping the performances represents a challenge for the next developments of the chip.}

The amplitude range of the \current{}  signal depends on the depletion of the detector. When under-depleted, part of the charges produced at the lowest energies are created in a region with null electric field, leading to a small current peak. From the data of ref.\cite{psa}, if we extrapolate our measurement to the lowest energies (10 fC) the smallest  amplitude goes down to 10~$\eta$A for under-depleted regime and 300~$\eta$A for over-depleted regime. The maximum amplitude (for 2200 fC) is extrapolated for the over-depleted regime at 100~$\micro$A for Z=1 particles and up to 200~$\micro$A for $^{12}$C. \added{The gain of the \current{} output was set at 7000~V/A, yielding a 1.4~V swing for a 200~$\micro$A input pulse.}

\added{Nevertheless the integration due to the parasitic components (detector capacitances, bonding wire, Kapton ribbon) and to the electronics integration, the \current{} rise time is dominated by the collection time, and in particular it is expected to range from 4~ns to 60~ns, requiring a bandwidth of at least 90~MHz.} The pulse width spread was specified from previous measurements with the PACI preamplifier. 

For the specifications of the iPACI, the shortest event is assumed to be 20~ns long. On the other hand, the maximum pulse width was observed for the He isotopes, yielding 300~ns in the under-depleted regime and 80~ns for over-depleted regime. Wider signals are expected for Li and C isotopes. 

The circuit is designed to operate on both sides of the detector, i.e. on both polarities. Furthermore, inter-strip events lead to inverse polarity with respect to normal events on the ohmic side \cite{Due14} and have to be captured to be able to retrieve the information from the inter-strip events. 
Moreover, the GRIT detectors present a variety of strip lengths implying that the preamplifier has to accommodate a range of detector capacitances from 10~pF to 40~pF including kaptons capacitance. 

Regarding heat management, the high compacity of the array allows only for 1 to 2~kW of power to be dissipated over the whole detector. Considering a system with approximately 7,500 detector channels, this yields a power consumption of 133~mW/detector channel, to be shared between the various circuits of the detector. For the preamplifier chip, the specification is 40~mW/detector channel.
The main specification items are summarized in table \ref{TopSpec}.

\begin{table}
\begin{center}
\begin{tabularx}{0.95\linewidth}{|X|X|} 
\hline
\multicolumn{2}{|l|}{\textbf{Outputs}} \\ \hline
\textbf{Energy max}									& 50~MeV in Silicon \\ 
\textbf{Bipolar}									& True \\ 
\textbf{ENC (Equivalent noise charge)}				& 10~keV FWHM (After optimal shaper, $C_{d}$=20~pF)\\ 
\textbf{Charge non-linearity } 					& $<2\%$ (up to 50~MeV)  \\ 
\textbf{Charge output recovery time} 				& $100 \micro$s  \\ 
\textbf{Current signal BW} 							& 90~MHz \\ \hline

\multicolumn{2}{|l|}{\textbf{System data}}  \\ \hline
\textbf{Technology} 								& AMS 0.35~$\micro$m BiCMOS \\
\textbf{Supply} 									& 3.3~V \\ 
\textbf{Detector's input capacitance} 				& Compatible with [10~pF .. 40~pF] range \\ 
\textbf{Current consumption} 						& $<40$~mW/detector channel\\
\hline

\end{tabularx}
\caption{iPACI main specification items}
\label{TopSpec}
\end{center}
\end{table}

\subsection{Architecture and circuit design}
\label{sec_archi}
\paragraph{Architecture}

The chip is composed of nine independent channels, each comprising a local bias block, a Charge-Sensitive Amplifier with Current Output and two buffers. The channel input is connected to one detector's strip, while the two buffer's outputs are connected to high speed digitizers via pads.

\begin{figure}[h]
	\begin{center}
		\includegraphics[width=\linewidth]{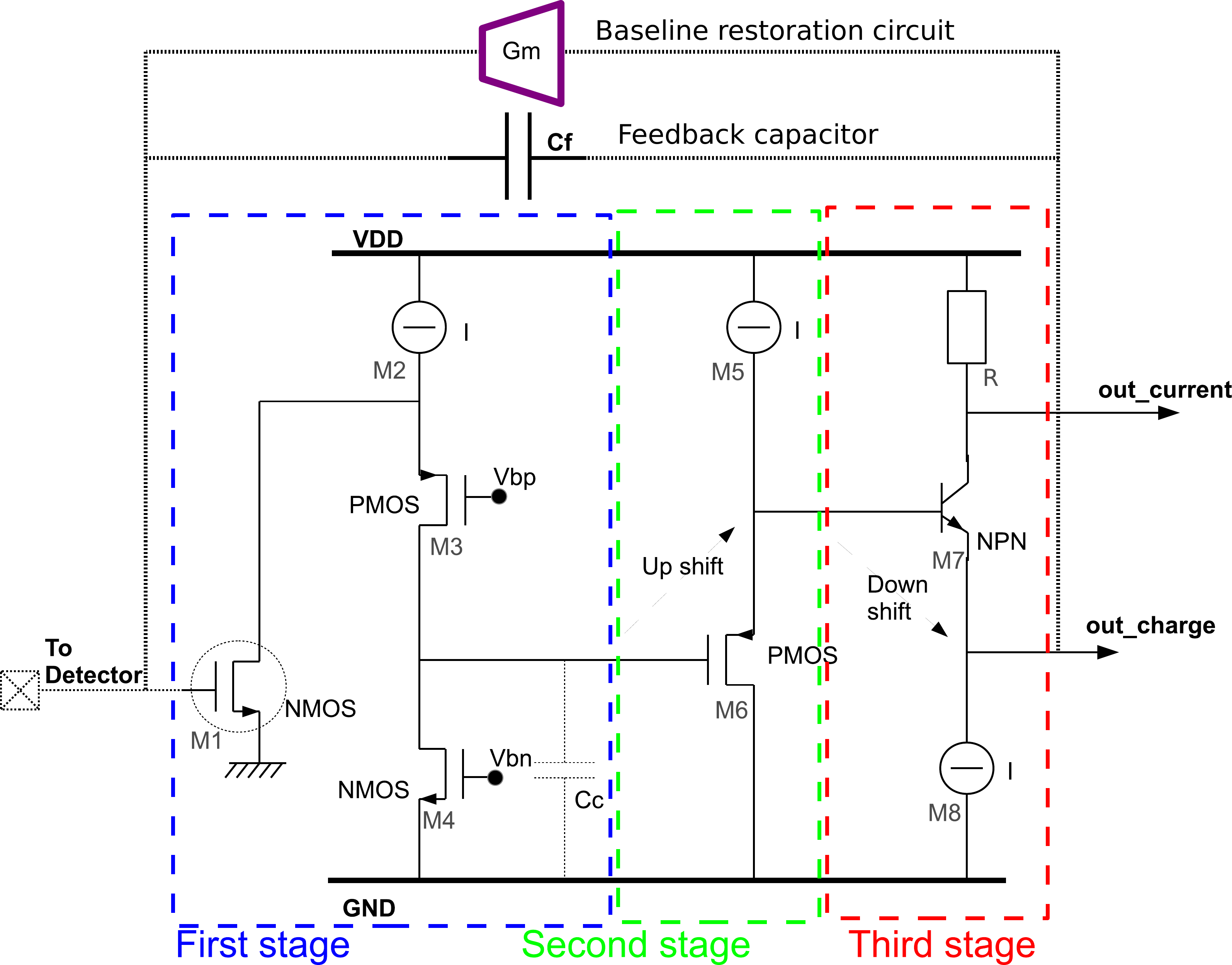} 
		\caption{Simplified iPACI architecture}
		\label{Simplified PACI architecture}
	\end{center}
\end{figure}

The architecture of one \CSACO channel is detailed in Fig. \ref{Simplified PACI architecture}. It is composed of an Operational Transconductance amplifier (OTA), a baseline-restoration amplifier (Gm) and a feedback capacitor ($C_f$).
With such an arrangement, the \CSACO acts as a current integrator, i.e. a virtual ground at the input absorbs all the current from the detector and generates a voltage output at the \outQ node, which is $V_{\textrm{out}} \approx I_d \cdot \frac{R_f}{1 + j \cdot R_f \cdot C_f \cdot \omega}$ (with $I_d$ the detector current, $R_f$ and $C_f$ respectively the feedback circuit's equivalent resistance and feedback capacitor). The baseline-restoration amplifier makes sure that the \outQ node returns to a target voltage within a given time after an event. This duration is typically of the order of 100~$\micro$s.

The \charge{} and \current{} outputs are single-ended continuous-time analog signals. The \current{} signal is the amplified image of the current received from the detector, while the \charge{} signal is the integral of the \current{} signal. Because a feedback resistance would need to be of several \added{M$\Omega$} and would be very large on chip, a MOS-based block ($Gm$), is used instead.

\paragraph{Operation}
The first stage is where all the voltage gain takes place. This gain can be rather large (around 60~dB) due to the folded cascoding of M1 by M3 and due to the large transconductance of the input device ($g_{m1}$) required for noise purposes. On the other hand, instabilities will occur here, due to the second pole, located at the source of transistor M3. A compensation capacitor, noted $C_c$, will need to be sized accordingly.

The second stage operates a voltage shift up by one Gate-Source voltage. It needs to be fast enough to pass the signal on to the next stage, which is easily achieved on a unity-gain stage.

The third stage operates a shift down by approximately the same amount of voltage and provides a low impedance drive for the \outQ signal. \added{The resistor} $R$ senses the current through the output branch, thus generating the \outI output; while M7 acts as a cascode that separates the \outQ and \outI parts of the branch. The required speed is also very easy to reach on this branch, due to the unity-gain operation of M7 and the use of an NPN transistor, which features higher transconductance than an NMOS.

\paragraph{Stability adjustment} 
In order to keep the preamplifier stable despite varying strip lengths, i.e. varying strip capacitances, the compensation capacitor ($C_c$) was made digitally adjustable across 8 compensation levels, i.e. 3~bits.

\paragraph{Baseline-restoration circuit}
The baseline-restoration circuit (Gm) insures that the preamplifier \charge{} signal returns to its base line after an event. It can be done with a simple resistor or a more complex device. In our case, a resistor would simply take too much area on silicon due to its high value. Also, the fact that the preamplifier has be be bipolar imposes a more complex solution.
A differential pair biased at 20~nA was used, which has the advantage of enabling the adjustment of the baseline voltage.
On the other hand, the differential pair acts as a current source for large signals, hence discharging the feedback capacitor at a constant 20~nA current. This, as a consequence, leads to a linear-shaped return the baseline, thus preventing to use pole-zero compensation, also leading to sub-optimal shaping.

\paragraph{Output buffer}
The output buffer is a two-stage miller-compensated single-ended OTA, wired up in a unity-gain configuration. This makes a simple design, as the OTA does not have to drive a low-resistance feedback circuit. Also, a unity gain insures the best speed.


\subsection{Simulated performance}
\label{sec_simu}
\added{As a first step, the iPACI prototype has been connected to a “simulated detector” modelled by a 20~pF strip capacitance, 5~uA total detector leakage current, equally spread across the strips, and biased by a 10~M$\Omega$ strip bias resistance. All simulations are done with the Spectre simulator \cite{spectre}, on Cadence Virtuoso 6.1.7, and using the AMS kit v.4.10. Note that unless otherwise specified, all figures were obtained at ambient temperature, 3.3~V supply, typical corner, maximum compensation capacitor setting, and with a 5~pF load on the output nodes, mimicking a PCB trace connected to a discrete operation amplifier. The performance achieved in this simulation are summarized in table~\ref{TopPerfs}. They show the feasibility of this project with reasonable consumption and area. }

 \begin{table}
\begin{center}
\begin{tabularx}{0.95\linewidth}{|X|X|} 
\hline
\multicolumn{2}{|l|}{\textbf{Charge Output}} \\ \hline
\textbf{Charge signal swing } 						& $\pm$1.6~V single-ended at 50~MeV\\ 
\textbf{Charge gain} 								& 32~mV/MeV \\ 
\textbf{ENC }										& 10~keV FWHM \\ 
\textbf{Charge output recovery time} 				& 100~$\micro$s  \\ \hline

\multicolumn{2}{|l|}{\textbf{Current Output \CSACO }}  \\ \hline
\textbf{Current gain} 								& 7~k$\Omega$ \\ 
\textbf{Current signal swing} 						& $\pm 0.750$~V single ended \\ 
\textbf{Current signal BW} 							& [4~MHz .. 130~MHz] \\ \hline

\multicolumn{2}{|l|}{\textbf{System data}}  \\ \hline
\textbf{Compensation capacitor } & Digitally tuneable [0.5~pF .. 2.25~pF], 0.25~pF step\\ 
\textbf{Current consumption} 						& 12~mA (40~mW)/ detector channel \\
\textbf{Area used on silicon (pads excluded)}		& 370$\times$200~$\micro$m/channel \\
\hline

\end{tabularx}
\caption{Simulated performance of iPACI.}
\label{TopPerfs}
\end{center}
\end{table}

\begin{figure}[h!] 
\begin{center} 
\includegraphics[width=.7\linewidth]{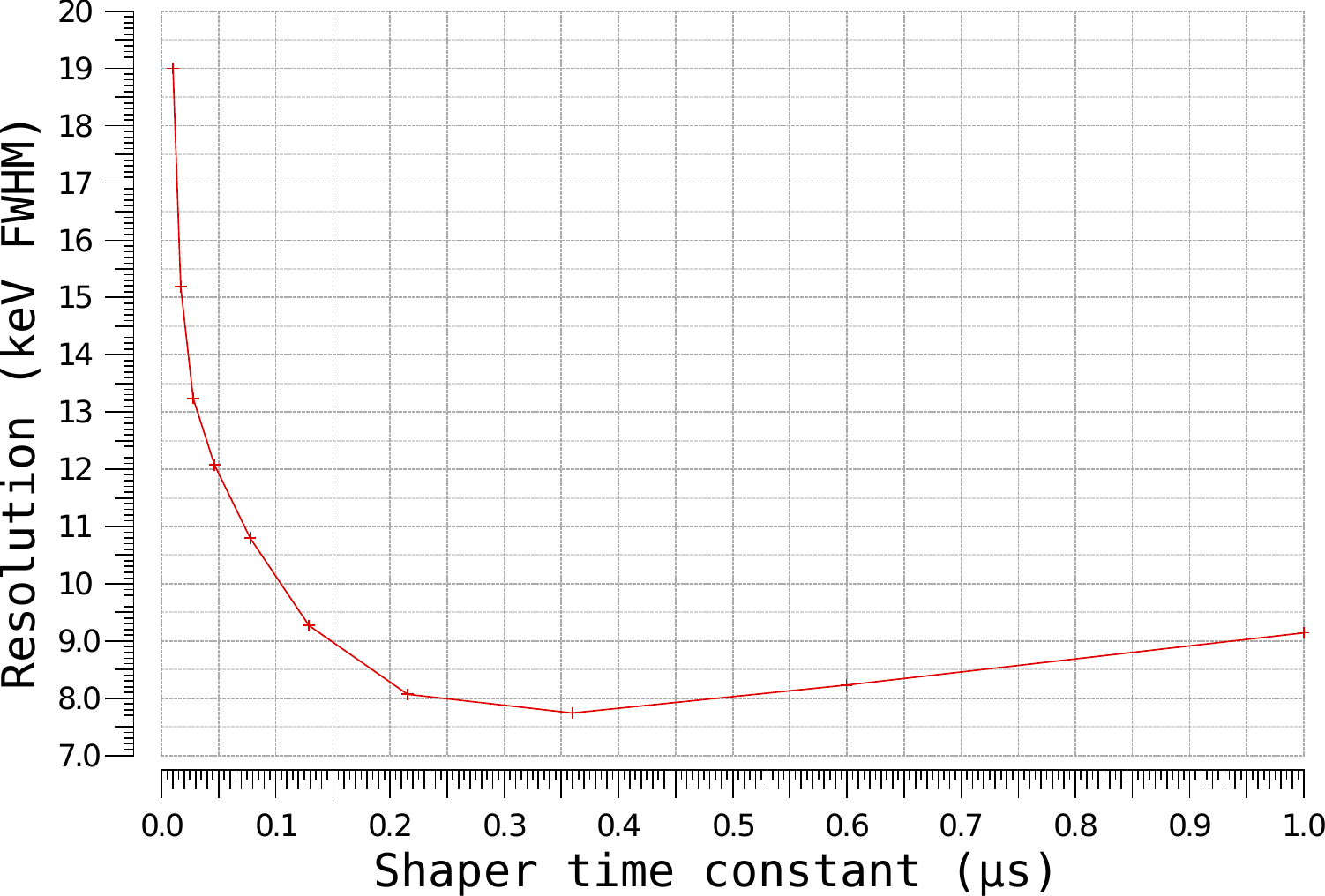} 
\caption{ENC vs shaper time constant} 	\label{ENCvsTau_sim} 	
\end{center} 	
\end{figure}

\added{In the following, we analyse the energy resolution connecting an ideal CR-RC2 shaper with adjustable time constant at the chip’s \charge{} output}. The noise measured at the shaper's output, in $V_{RMS}$ is referred back to the input in keV FWHM via the input-to-output transient gain, found via a transient simulation.-
Fig.~\ref{ENCvsTau_sim} shows the resolution obtained versus the shaper time constant. The optimum is found at 7.7~keV at 360~ns.
Table~\ref{TopNoiseContributors} presents the main contributors to the noise figure obtained on the previous simulation at the shaper's optimal time constant (360~ns). 
\begin{table}[h!]
\begin{center}
\begin{tabular}{|l|c|}
\hline 
\textbf{Contributor}  & \textbf{Percentage} \\ 
\hline 
\textbf{Current bias circuit} & 32.34 \\ 
\textbf{\CSACO}          & 13.87 \\ 
\textbf{Detector}     & 24.02 \\ 
\textbf{Baseline restoration} (Gm) & 26.69 \\ 
\textbf{Various}     & 3.08 \\
\hline 
\end{tabular} 
\caption{Main noise contributors}
\label{TopNoiseContributors}
\end{center}
\end{table}
\added{One notices a predominance of the \textit{bias circuitry}, used to bias the analog stages at well defined current and very critical in low-noise designs, as largest contribution to the noise}. The \textit{detector} also contributes a significant amount of noise via its intrinsic noise ($2q \cdot I_{leak}$) and the strip bias resistance. Finally, the \textit{baseline-restoration circuit} has a significant impact, as it is directly connected to the input node, and constantly biased. As a conclusion, the table shows a quite balanced design, with all main contributors having a similar share.

\begin{figure}[h!] 
\begin{center} 
\includegraphics[width=.8\linewidth]{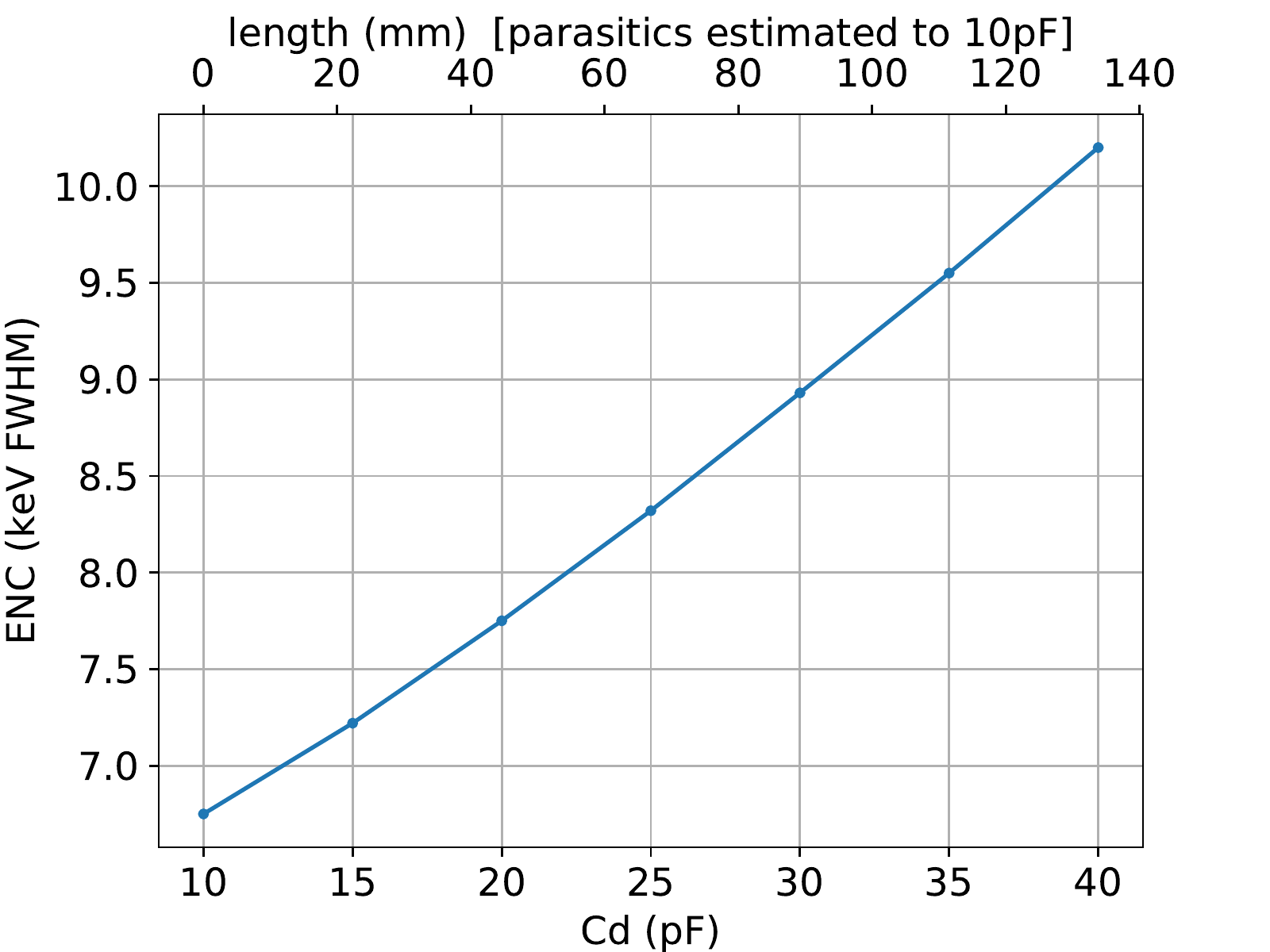} 
\caption{ENC vs Detector capacitance simulated with a 360~ns shaping time. Note that we estimated the parasitic capacitance (brought from the PCB traces) to 10~pF.}
\label{ENCvsCd} 	
\end{center} 	
\end{figure}

The simulation of the resolution versus the detector capacitance (Fig.~\ref{ENCvsCd}) shows the effect of the strip length, directly related to the strip capacitance.

\subsection{Test results}
\label{sec_test}


\begin{figure}[h] 
\begin{center} 
\includegraphics[width=0.8\linewidth]{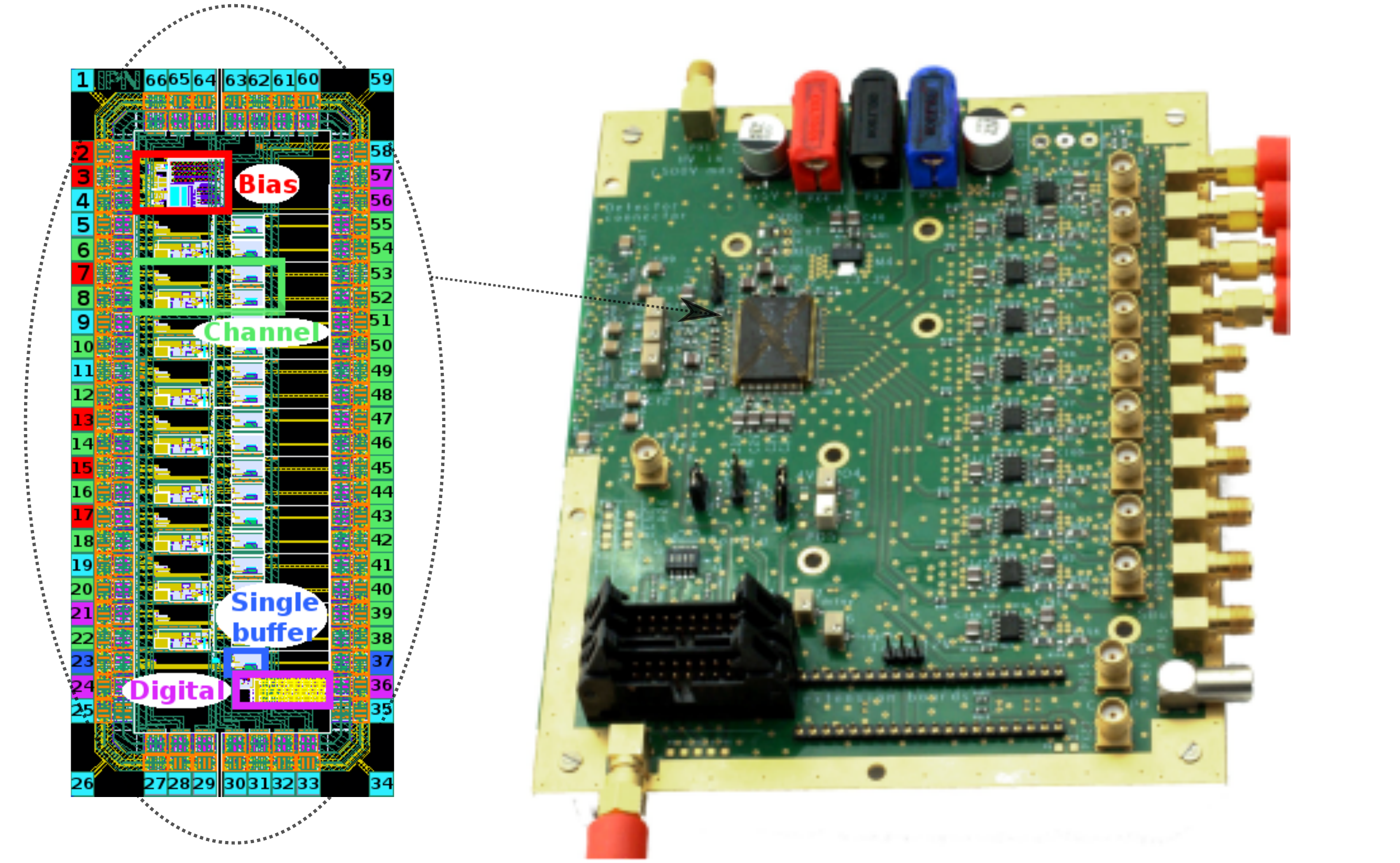} 
\caption{IPACI chip and Test board} 	\label{iPACI_chip} 	
\end{center} 	
\end{figure}

The ASIC was produced in AMS 0.35~$\micro$m BiCMOS SiGe technology. In order to test the ASIC, a test board was designed and is shown on Fig.~\ref{iPACI_chip}. The iPACI test board embeds one iPACI, connected to a detector connector and injection circuitry at its input; and to discrete output buffers at its output, driving 50~$\Ohm$-terminated coax cables. Unless otherwise specified, all figures were obtained at ambient temperature, 3.4~V supply and maximum compensation capacitor setting.
In this section, we present the test results and issues seen with this electronics.

\begin{figure}[h]
	\begin{center}
		\includegraphics[width=\linewidth]{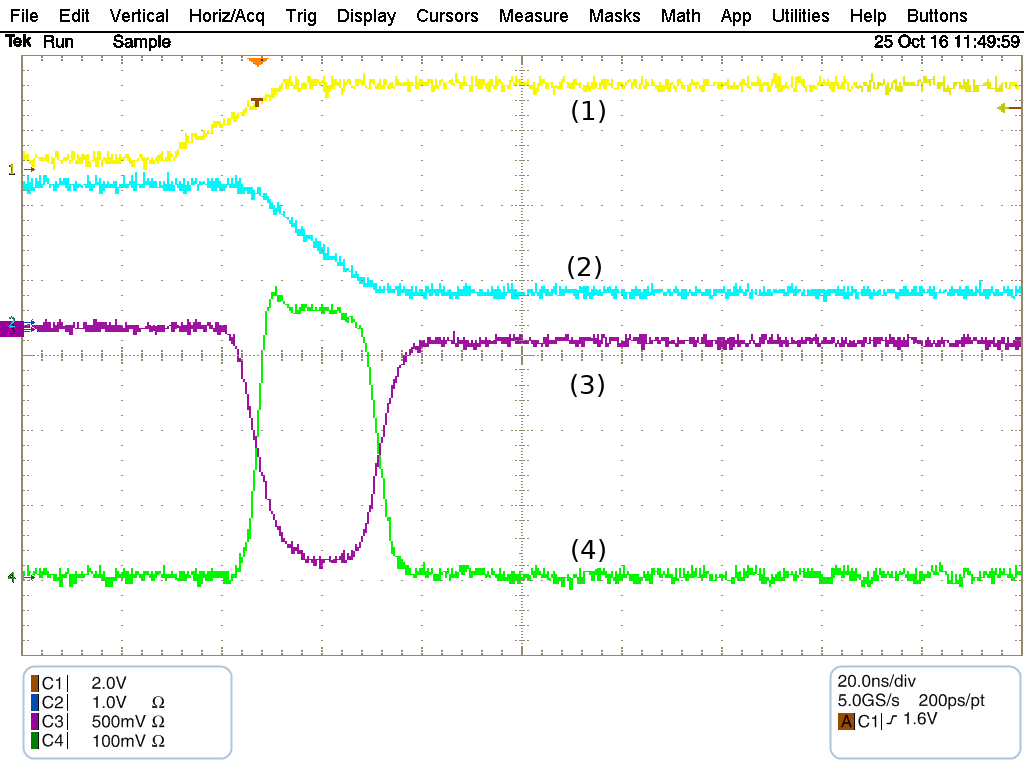} 
		\caption{Injection and output signals. (1) 50~Mev injection in 25~ns (2) \charge{} signal (3) iPACI \current{} signal (4) Alternate current signal, discussed in the Stability part.}
		\label{ExtDiff}
	\end{center}
\end{figure}

\paragraph{Injection} An injection signal of amplitude equivalent to 50~MeV was sent to the chip in order to verify that the functionality was correct. 
For that purpose, a function generator is connected to the detector input via a 1~pF capacitor, and is programmed to generate ramps of 25~ns rise time with adjustable amplitudes. The amplitude is calculated to match the required charge. The \charge{} and \current{} signals are in turn digitized by an oscilloscope (Tektronix TDS7154B). They can be seen Fig.~\ref{ExtDiff}, showing correct functionality.

\begin{figure}[h]
	\begin{center}
		\includegraphics[width=\linewidth]{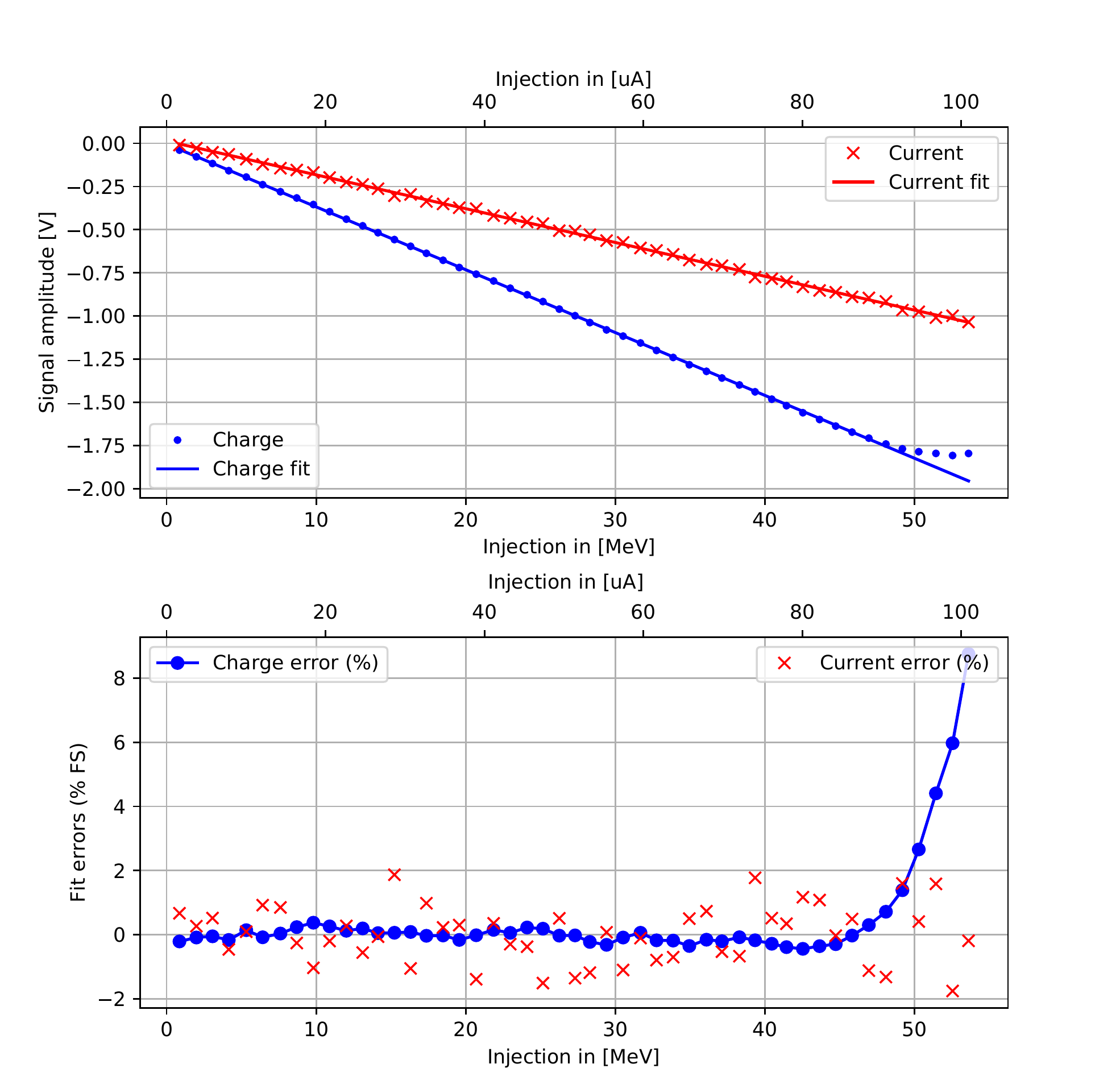} 
		\caption{Linearity plot. (TOP) \charge{} and \current{} output voltages (V) as a function of the injection amplitude (shown both in~MeV and~$\micro$A)  (BOTTOM) Deviation from the fitted curve, normalized in \% of the full scale.}
		\label{Linearity}
	\end{center}
\end{figure}

\paragraph{Linearity} In order to measure the linearity, energies of increasing values are sent to an ASIC channel, while the output \charge{} and \current{} signals are digitized \added{with the same setup as previously} and fed to a PC. \added{\textit{Charge} and \current{} peak voltage values are reported (Fig.~\ref{Linearity}, TOP) as a function of the input energy, expressed in~MeV and~$\micro$A. The data are fitted using a linear regression, yielding a slope value for both measurements. For each data point measured on the charge output, the voltage deviation between the actual measurement and the fitted curve is referred back to MeV using the slope calculated previously. The results are plotted Fig.~\ref{Linearity} (BOTTOM), as a percentage of the full scale (50~MeV), yielding a maximum non linearity lower than 2\% up to 49.7~MeV injection, and rising very quickly after due to saturation.}
The current measurement is processed in the same way, yielding 2\% maximum non linearity.

\paragraph{Speed} The speed performance of a preamplifier can easily be measured using the \current{} output, which is an image of the input current, however limited by the system's bandwidth. Speed can either be quoted in terms of bandwidth and measured via a small signal AC injection; or in terms of rise time, and measured as a result of a fast pulse input. We measured the iPACI speed performance using both methods.

In order to run the AC test, a programmable sinewave generator (Rohde\&Schwarz SMX) is connected via a resistor to the channel input, and programmed to sweep its frequency from 10~MHz to 250~MHz. 
The output sinewave seen on the \current{} output is digitized by an oscilloscope, and the traces are sent to a PC, which measures their peak-peak amplitudes. From the data, we measure 130~MHz (at 3dB).

The transient test is more straightforward. A high-performance pulse generator injects fast voltage ramps via a 1~pF capacitor. At 50~MeV equivalent injection, we measure a rise time of 4.3~ns (10-90\%) on the \current{} signal.

Note that the $4.3~ns$ figure leads to a large-signal bandwidth of approximately 80~MHz, versus a small signal bandwidth of 130~MHz. This discrepancy is caused by the A-class nature of the stages used in the miller amplifier. This yields to a slew rate which is limited by the stages bias currents, i.e. a low large-signal bandwidth.

\paragraph{Equivalent noise charge (ENC)} The ENC is a measure of the noise present on the \charge{} signal after an ideal shaper. The ENC is expressed in~keV FWHM, hence requires to refer the output noise that is measured back to the input. For that purpose, the input-to-output gain is measured on a transient measurement of an energy-calibrated event, like it was done on simulation.

The noise measurement is done in steady-state after a low-noise shaper and an RMS-voltmeter (specific hardware was used), in order to simplify the setup. The shaper's time constant is adjustable, and was swept across its whole range (10~ns to 1~$\micro$s).

Several test setups were measured in order to fully qualify the system. As a first step, the shaper was measured on its own in order to make sure that its contribution to the overall noise was negligible. Then, the preamplifier and shaper only in order to qualify the noise contributed by the electronics. A PCB fitted with 20~pF capacitors was then plugged onto the detector connectors, thus enabling measurement of the detector's parasitic capacitance contribution to the ENC. Finally, a DSSD was put in place and biased to its nominal voltage, showing the contribution of both parasitic capacitance and leakage current. \added{The DSSD used is extensively described paragraph \ref{sec_detector}, followed by in-beam test results. }

\begin{figure}[h]
	\begin{center}
		\includegraphics[width=\linewidth]{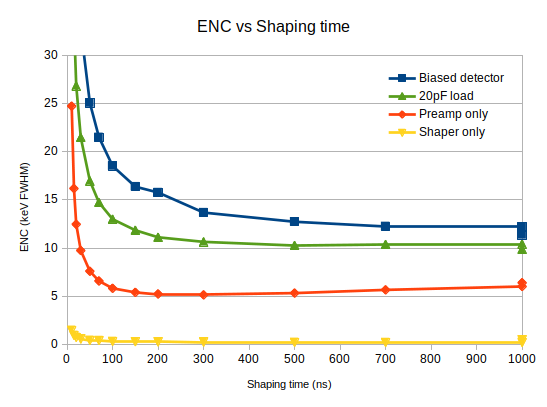} 
		\caption{Equivalent Noise Charge measured on the following conditions: \textbf{BIASED DETECTOR} Measurement done with the detector connected and biased to its nominal depletion voltage; \textbf{20pF LOAD} 20~pF capacitive loads (equivalent to 44~mm strip) are connected to the preamplifier channels, thus emulating a detector; \textbf{PREAMP ONLY} The preamplifier has its input disconnected, thus showing the contribution of the electronics to the ENC; \textbf{SHAPER ONLY} The preamplifier is disconnected, showing the shaper's contribution to the ENC}
		\label{ENC}
	\end{center}
\end{figure}

The results are shown Fig.~\ref{ENC}, showing 12~keV FWHM ENC at 850~ns in presence of the detector.
The measured value is very close to the simulated one. The discrepancy can easily be explained by the non idealities on the PCB (bond wires, supply decoupling, EMC) and unmodeled  non-idealities of the detector

\begin{figure}[h]
	\begin{center}
		\includegraphics[width=\linewidth]{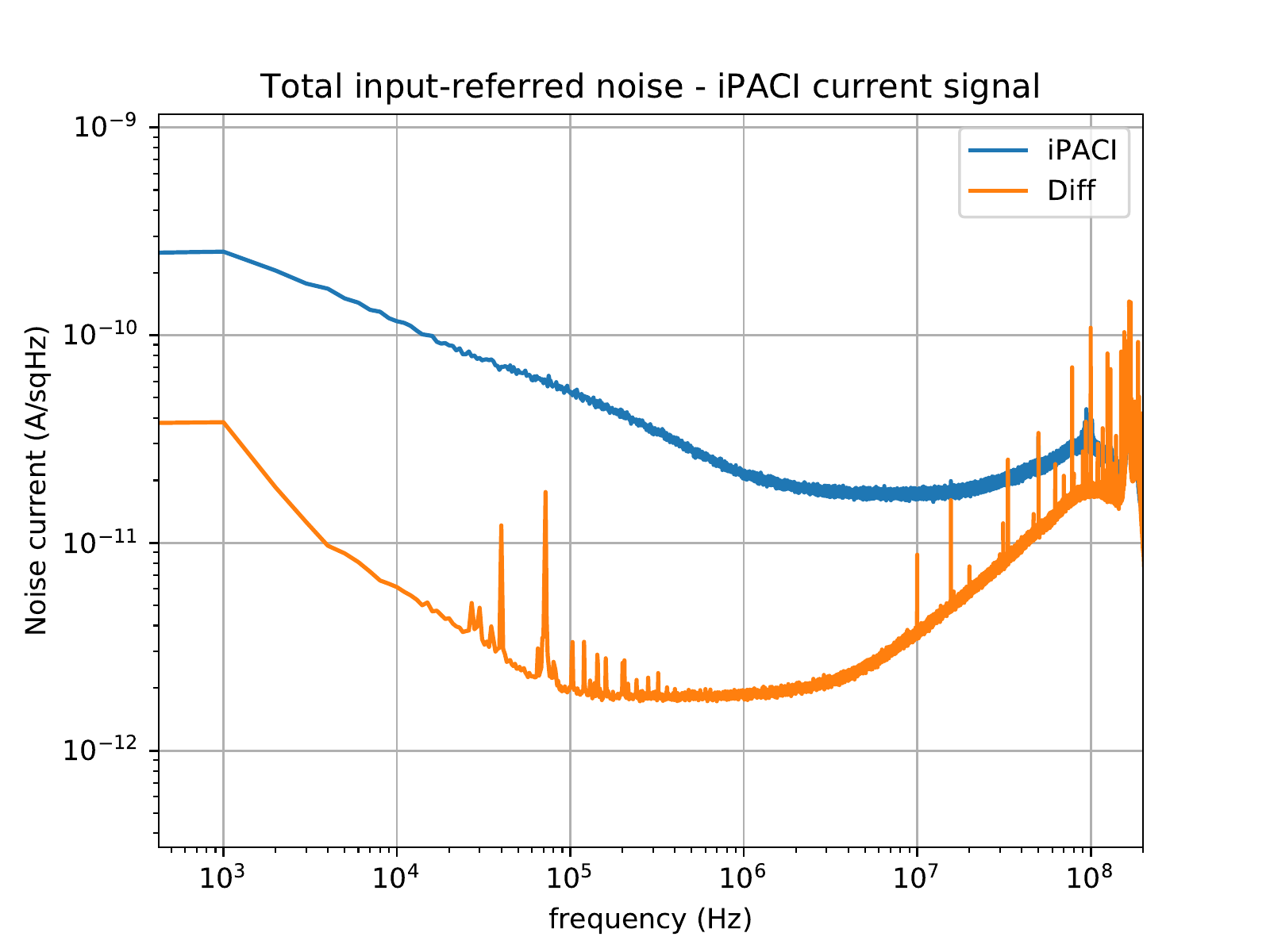} 
		\caption{Equivalent Noise current. (IPACI) iPACI's native \current{} output. (DIFF) Differentiator's output. }
		\label{fig:ENCu}
	\end{center}
\end{figure}

\paragraph{Equivalent noise current} In order to measure the equivalent noise current, we measure the noise voltage on the \current{} output and refer it back to the input using the gain from input to output. Similarly to the ENC, we use the gain seen on a calibrated event.
More specifically, a low-noise 100-gain voltage amplifier is inserted between the \current{} output and an oscilloscope. Noise traces were recorded and fed to a PC, which processes and plots the periodograms. In order to reduce the noise on the plot, 32 successive periodograms are averaged together.
The obtained plots are shown in  Fig. \ref{fig:ENCu}. They were made on the native iPACI's \current{} output (labelled iPACI on the figure) as well as on the differentiator's output (labelled Diff) (see section \ref{Issues} for details on the differentiator circuit). They suggest that the native \current{} output noise could be greatly improved with a redesign, thus reaching the performance of the differentiator circuit.

\label{Issues}

\label{stability}
\paragraph{Stability} Once a silicon detector is connected to the ASIC inputs, it strongly destabilizes the preamplifier. This effect is due to the strong strip capacitance seen from the preamp's input to the experiment's ground\added{, and can be reproduced with the detector emulator PCB described earlier}. This creates a strong sensitivity to the ASIC's internal ground noise, which gets created in particular during output spikes.
%

\begin{figure}[h]
	\begin{center}
		\includegraphics[width=\linewidth]{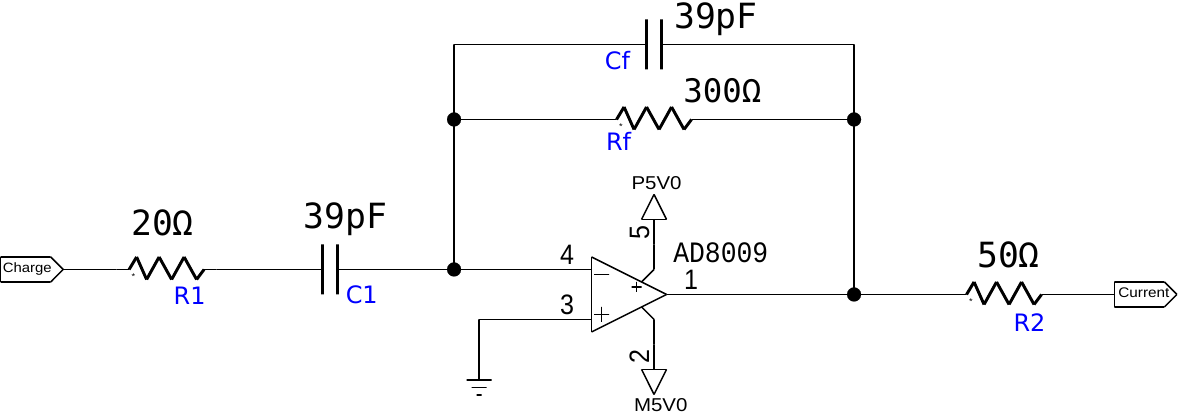} 
		\caption{Circuit used for differentiation}
		\label{Diffcircuit}
	\end{center}
\end{figure}

Two ways were found to mitigate the issue, which are shown in Fig. \ref{ExtDiff} (signals 3 and 4).
The first is to reduce the current spikes coming from the \current{} output buffer by inserting a resistor of small value on its output, however limiting the bandwidth of the \current{} signal.
The other one is to differentiate the \charge{} signal on PCB, using commercial opamps, at the expense of higher PCB area usage and significant additional consumption (Fig. \ref{Diffcircuit}). 

In this configuration, the opamp creates a virtual ground on its negative input. R1 feeds a current to the feedback network that is a derivative of the input signal. Rf converts this current to a voltage, also enabling adjustment of the conversion gain. R1 and Cf are present only to slightly reduce the bandwidth of the differentiator circuit, thus preventing instabilities.


\section{Results of the first in beam test of the iPACI prototype with the GRIT trapezoidal prototype DSSD detectors}
\label{sec_beam}

\subsection{Trapezoidal DSSD from GRIT}
\label{sec_detector}

The iPACI preamplifier was tested with a GRIT prototype detector in beam. For this test, we used a 375~$\micro$m thick trapezoidal DSSD of nTD type with a cut at 4 degrees (crystal orientation $<$100$>$) to avoid channeling \cite{Bard09_2}. The DSSDs are manufactured by Micron Semiconductors Limited \cite{micron} in 6{\nbd}inch technology. 
The design is custom-made (see Fig. \ref{trap}) with the characteristics detailed in table \ref{tabSi}. The packaging of the detector is designed to get as narrow frames as possible and the 20~mm long kapton ribbon cables are bent at 80 degrees and ended by two high density 80-pins connectors. The strips on the ohmic side are perpendicular to the strips on the junction side. The strip pitch is 800~$\micro$m on the ohmic side and 700~$\micro$m on the junction side. All its characteristics will be the final one except the thickness of the detector that will be 500~$\micro$m for the final detectors.
The detector was reverse-mounted for the experiment i.e. the ohmic side was facing the beam. Its nominal bias according to the constructor specification is 120~V. At this bias, the energy resolution from the alpha source calibration was found to range from 28 to 36~keV on the N-side and from 32 to 40~keV on the junction side. 

\begin{table}
\begin{center}
\begin{tabular}{|c|c|}
\hline \textbf{Characteristic} &\textbf{FFF2}\\
\hline
Wafer type & nTD 6 inch\\
Strip pitch ohmic side& 760~$\micro$m\\
 Strip pitch junction side & 710~$\micro$m\\
Crystal orientation & $<$100$>$\\
Resisitivity & 2000 $\Omega$.cm\\
Thickness & 375~$\micro$m\\
Active area & 5000~mm$^{2}$\\
Height & 105~mm\\
Metal coverage & Al 30~$\micro$m\\
Dead layer & $<$1~$\micro$m\\
Number of strips & 128X $+$ 128Y\\
\hline
\end{tabular}
\caption{Characteristics of the GRIT prototype detector from Micron Semiconductors Limited.}
\label{tabSi}
\end{center}
\end{table}

\subsection{Experimental set-up and electronics chain }

The in-beam test of the electronics and detector of GRIT  was performed at the Tandem-ALTO facility in Orsay with a $^7$Li beam at 35~MeV impinging on a 100~$\micro$g/cm$^2$ $^{12}$C target.
This reaction has been extensively used for earlier PSA studies \cite{psa,Due12} as it produces a large variety and amount of Z=1 and Z=2 particles. 

As the goal of the experiment is to focus on particle identification, the DSSD was under-depleted at 80~V \added{taking into account the reverse current} (the depletion being 90~V) for the experiment to improve particle identification \cite{psa}. The leakage current in this case was 0.7~$\micro$A corresponding to a 7~V drop in the applied bias and it was stable all along the experiment.

\begin{figure}[h]
	\begin{center}
		\includegraphics[width=\linewidth]{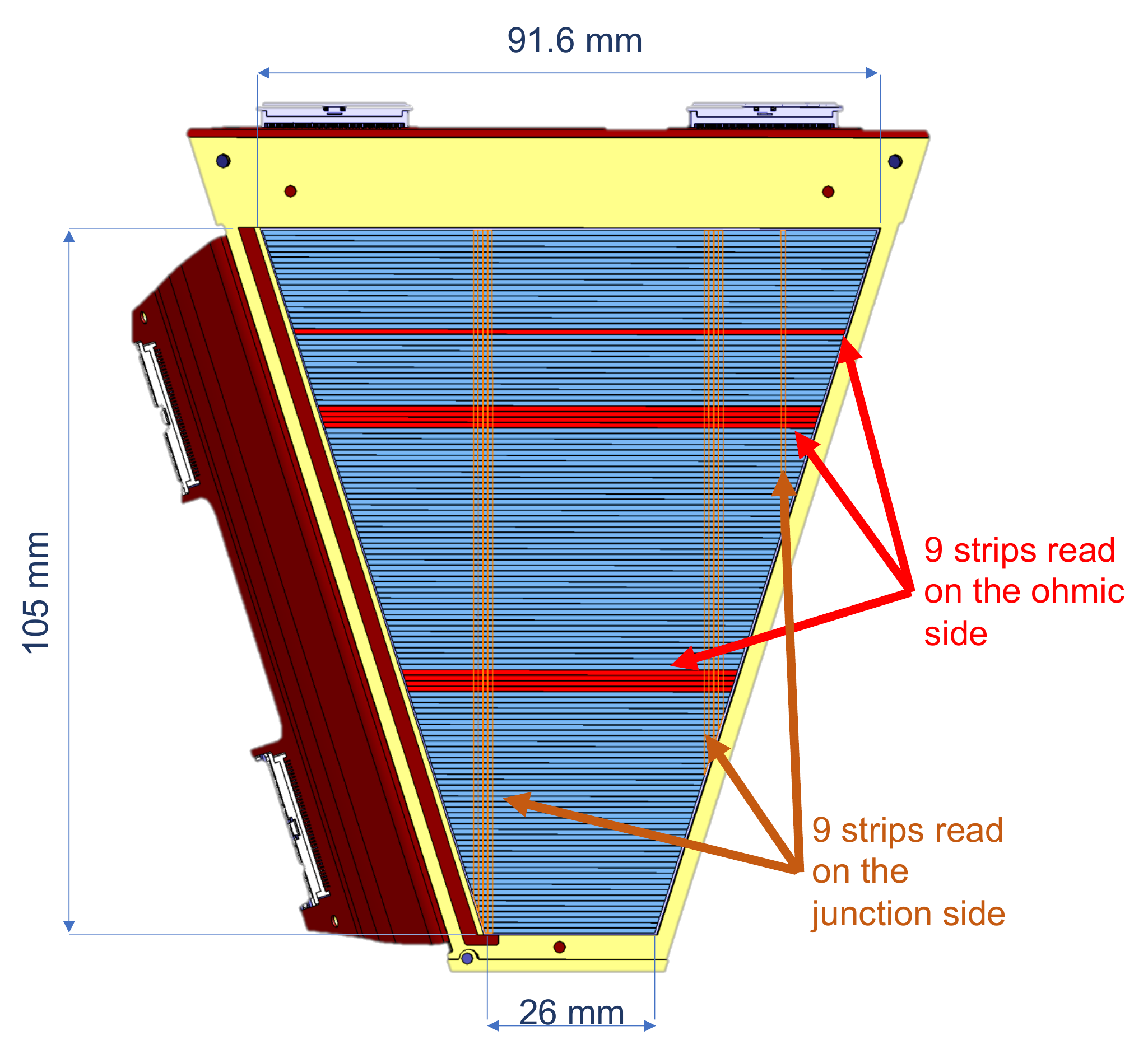} 
		\caption{3D drawing of the trapezoidal detector used in this experiment seen from the ohmic side. The 9 strips read on each side are shown in red for the ohmic side (N-side) and in orange for the junction side (P-side).}
		\label{trap}
	\end{center}
\end{figure}

Given the trapezoidal shape of the detector, the strips have different lengths depending on their position along the detector. This is why samples of neighbouring strips with different lengths were chosen to be read for the test (see Fig.~\ref{trap}). A total of 9 strips were read on each side of the detector with one 9-channel iPACI ASIC for each side providing for each channel the \charge{} and \current{} signals. The iPACI chips were sitting at 5~cm from the detectors, in vacuum and were cooled down to 15 degrees. The total of 36 outputs (\charge{} and \current{} signals for both sides) were read by the Wavecatcher digitizer \cite{wavecatcher,waveC1}, a 12-bit digitizer board with $\pm$1.25~V input dynamics. The Wavecatcher was set at a sampling rate of 400~MSa/s and a total window of 1024 samples with a trigger delay of 176 samples to cover the full bandwidth of both Silicon detectors and preamplifiers. The trigger was generated internally on \charge{} signals with a discriminator threshold of 10~mV. If one channel triggered, all the 36 signals were written on disk. The maximum acquisition rate for 36 channels was about 100~Hz. 

\subsection{Pulse shape discrimination results}

\subsubsection{Data processing}

\begin{figure}[h]
	\begin{center}
		\includegraphics[width=0.7\linewidth]{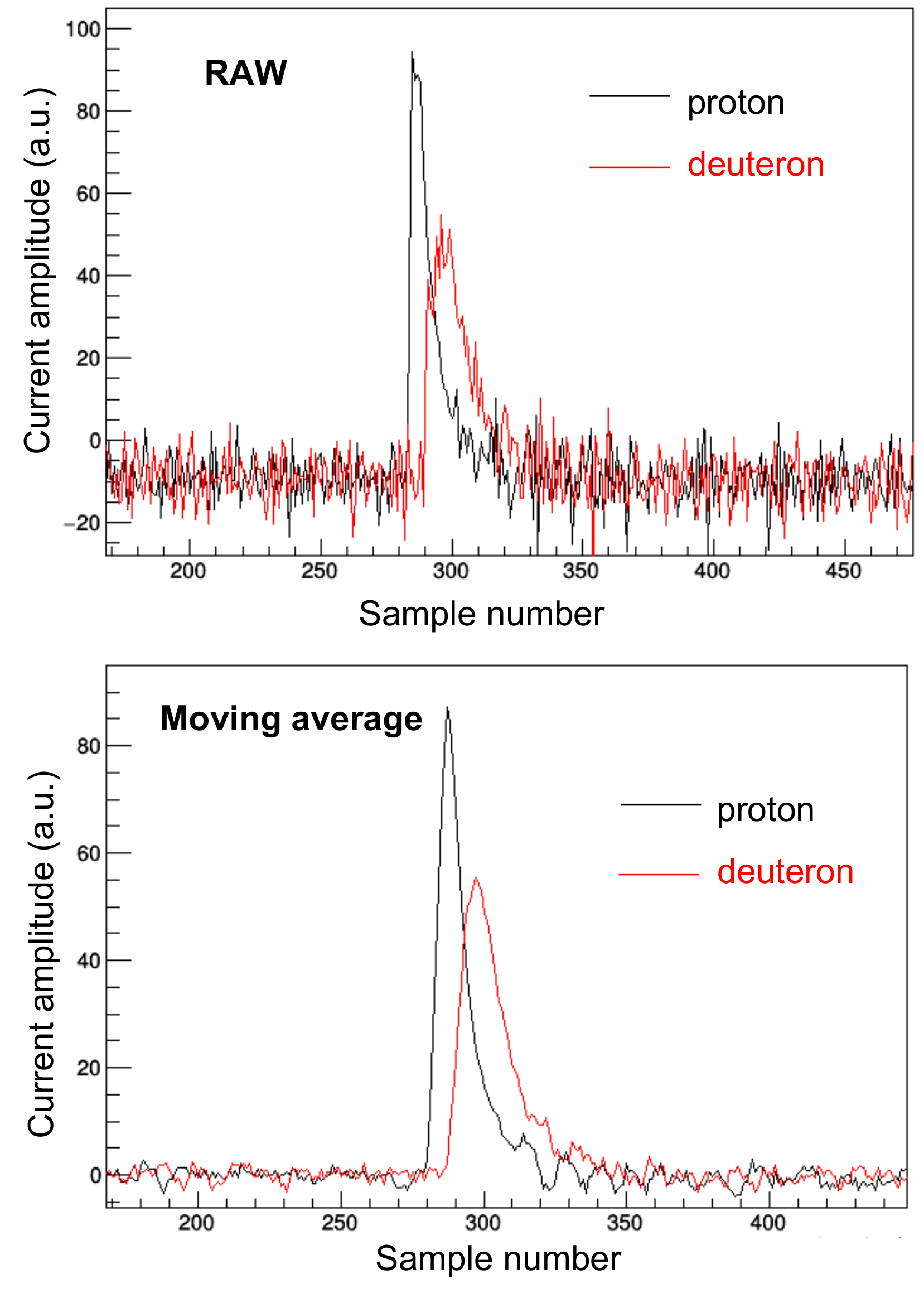} 
		\caption{Typical signals for proton (black) and deuteron (red) at 5~MeV. The raw signals are shown in the top panel whereas the filtered ones (using a moving average on 7 samples, see text) are shown at the bottom. }
		\label{signals}
	\end{center}
\end{figure}

The first step of data processing is to remove the baseline of both \charge{} and \current{} signals by averaging the first 176 samples (i.e. noise samples before the actual peak).
Typical raw \current{} signals for a proton and a deuteron of 5~MeV as digitized by the WaveCatcher are shown on Fig. \ref{signals} (top). 
These signals are then filtered by applying a moving average on 7 samples for the \current{} signal corresponding to a bandwidth of 25~MHz. The corresponding signals are shown on Fig. \ref{signals} (bottom). The amplitude of the current signal (Imax) is determined after the filter as the maximum of the signal.

The charge signal is filtered by applying a moving average on 101 samples (1.75~MHz bandwidth). The energy of the particles is deduced from the maximum of the \charge{} signal. The energy calibration was performed using triple alpha sources and the punchthrough of the Z=1 particles and of the alpha particles.

Interstrip events can be further suppressed by requiring the coincidence between the P- and N-side with the condition that the energy difference between both sides is less than 0.5~MeV.

\subsubsection{Pulse shape discrimination and effect of strip length}

\begin{figure*}[h]
	\begin{center}
		\includegraphics[width=0.9\linewidth]{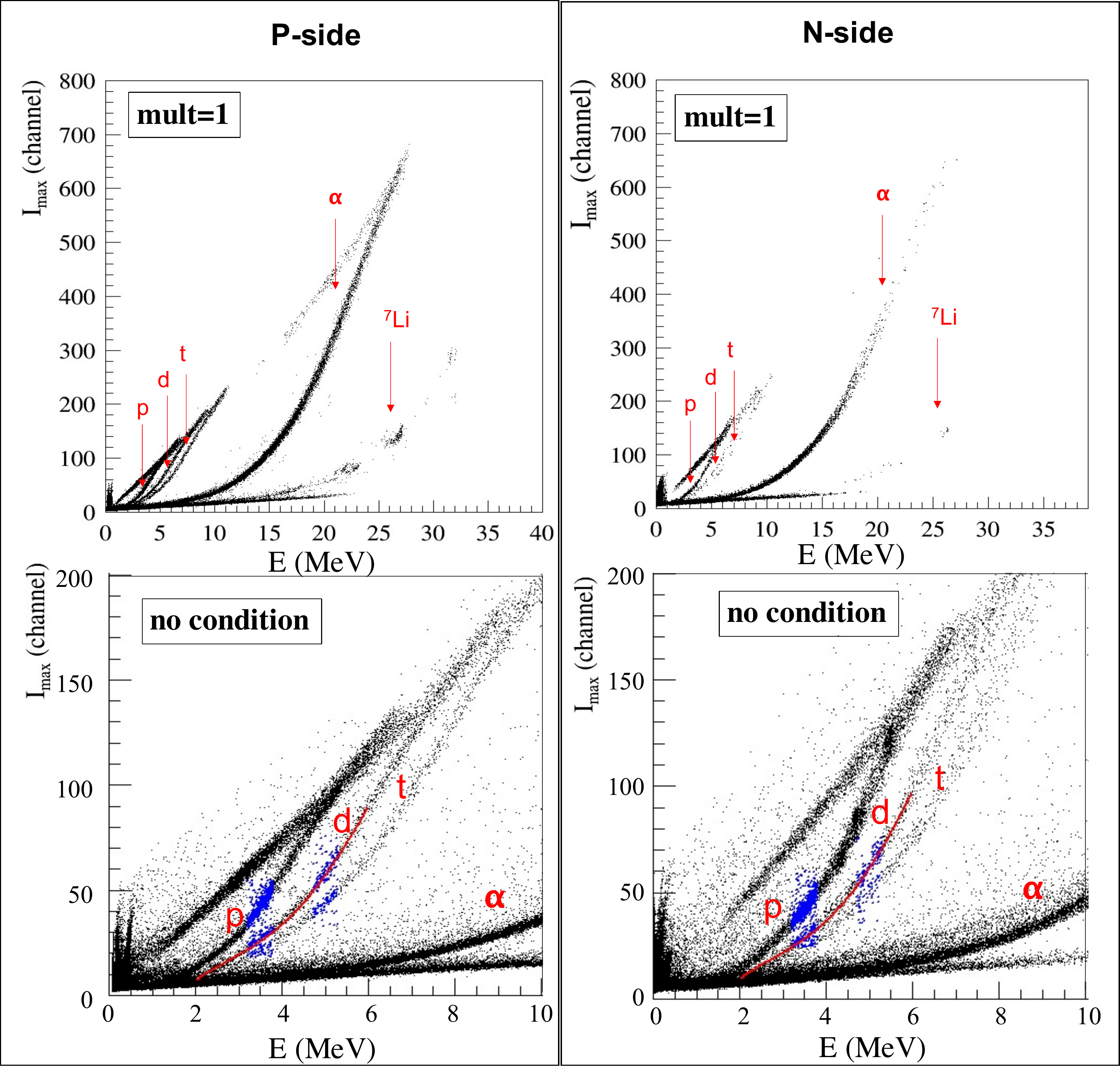} 
		\caption{Imax versus Energy spectrum for one N-side strip (left) and P-side strip (right). The top spectra present the full energy range with a condition on the multiplicity to be equal to 1. The bottom spectra are zoomed on the Z=1 isotopes.  In red is shown the fit of the deuteron line and in blue the selected events around 3.5~MeV for proton-deuteron separation and 5~MeV for deuteron-triton separation.}
		\label{IQ}
	\end{center}
\end{figure*}

The pulse shape discrimination technique relies on the fact that different particles with the same energy will have different interaction ranges in the detector. The so-called plasma delay time \cite{Pau94} increases with the mass and influences the rise time of the signals whereas the fall time is mainly attributed to the drift of charge carriers. On Fig.~\ref{signals}, the difference between the proton and deuteron signals at 5~MeV can be clearly seen: the height and the width of the signals are completely different although the integrals are equal.
For GRIT, the particle identification on both the N- and the P-side relies on the amplitude of the \current{} signal (Imax). The spectra obtained for one strip of each side are shown on Fig. \ref{IQ} (top) in the under-depleted case (80~V) \added{ with the condition that the multiplicity is equal to 1.} The signal from the protons, deuterons and tritons on one side and from $^4$He and then Li isotopes on the other side align on an almost parabolic shape. The punchthroughs of the protons, deuterons, tritons appear as a straight line and are mixed up. For the $^4$He, it is only observed for the N-side strip. Another line appears below the Li line and stops at 22~MeV. It is tentatively assigned to Be isotopes.

The lowest energy threshold for effective particle identification, determined as the separation between the different isotopes, is 2~MeV for protons, 2.5~MeV for deuterons and 3~MeV for tritons and He isotopes. Below these values, where the particles are very slow, the time-of-flight can be used to identify them. 
More challenging are the particles with the highest energies near the punchthrough points. For these particles, the rise time of the \current{} signal is very small. For example, it is 4~ns for a 6~MeV proton or a 25~MeV alpha. Given the 400~MSa/s sampling rate, the rise time is spread over very few samples. However, after filtering and thus reducing the bandwidth, the discrimination is still clearly achieved.

\begin{figure}[h]
	\begin{center}
		\includegraphics[width=0.9\linewidth]{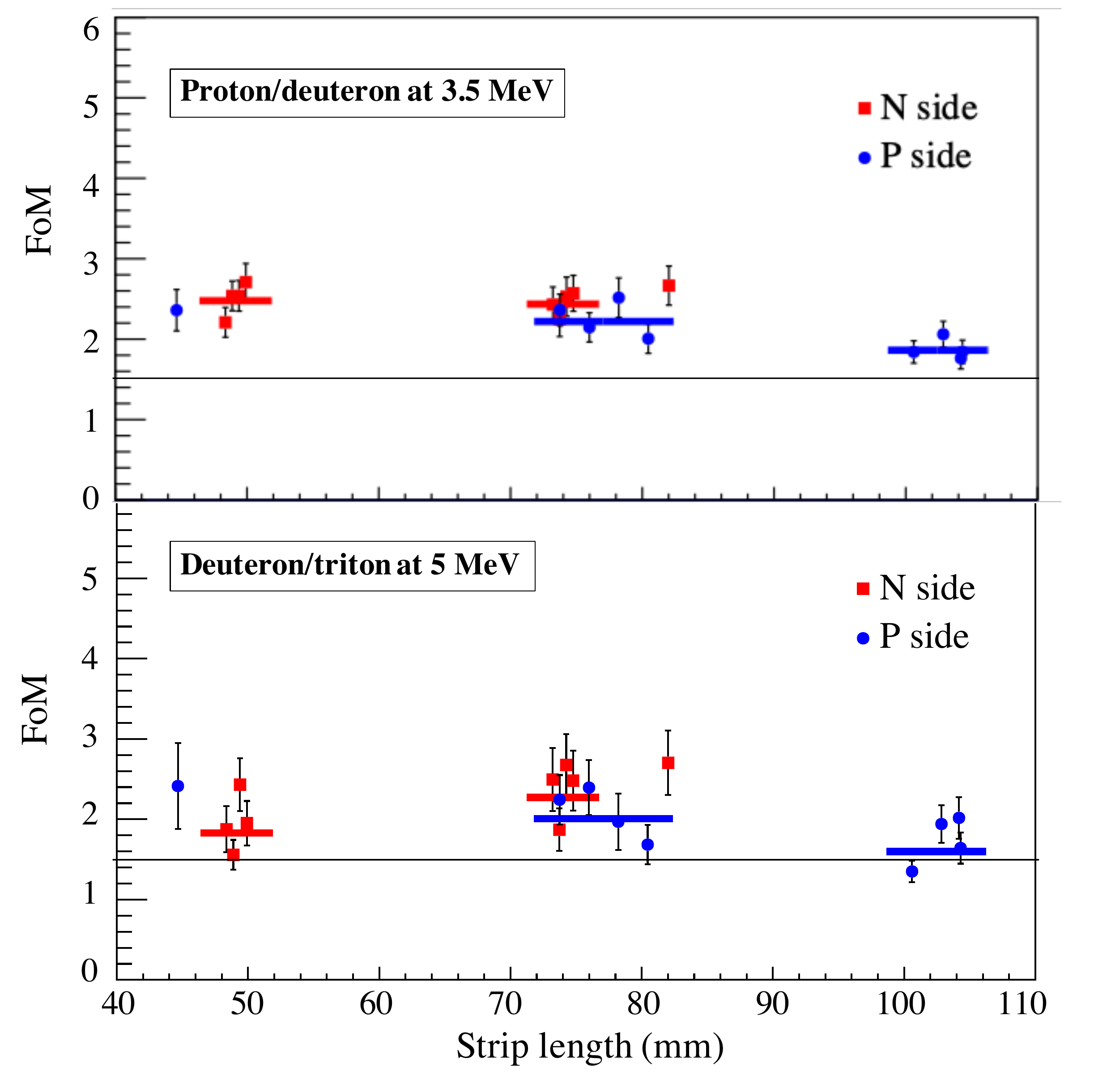} 
		\caption{Figures of merit as a function of strip length for proton deuteron separation at 3.5~MeV (top) and for deuteron triton separation at 5~MeV for under-depletion regime (80~V). In blue are shown the P-side strips and in red the N-side strips. The horizontal lines at 1.5 give the limit for particle discrimination.}
		\label{fom}
	\end{center}
\end{figure}

The figures of merit (FoM) for the separation between the Z=1 isotopes are evaluated using the following Particle Identification (PId) technique \added {on the spectra without any condition on the multiplicity to get more statistics}. A two-dimension fit with a first or higher-order polynomial  (depending on the observable considered) is performed on the "curve" corresponding to a particle (deuteron in our case) (for example, for the deuteron as shown in red on  Fig. \ref{IQ} (bottom)). Then, the PId corresponding to the minimum distance to the curve is calculated for the selected events shown in blue on Fig. \ref{IQ} (bottom).
In order to qualify the separation between the different isotopes, each PId component is fitted with gaussians and the figures of merit are calculated with the following definition: 
\begin{equation}
\textrm{FoM} = \frac{2|m_1-m_2|}{2.35(\sigma_1+\sigma_2)}
\end{equation}
where $m_1$ and $m_2$ are the peak positions and $\sigma_1$ and $\sigma_2$ are the standard deviations for each peak.  A FoM of 2 corresponds to a separation between the two peaks of the sum of their FWHM (in the gaussian approximation). The criterion for particle identification with this definition is: FoM$\geq$1.5.
The FoM cannot be calculated over the full energy range because the shape of the Z=1 isotopes lines are quite different. So we evaluated them at a fixed energy within a 500~keV wide gate, taking the deuteron line as a reference. To optimize the statistics for proton-deuteron separation, it is calculated at 3.5~MeV and for deuteron-triton at 5~MeV (see blue points on Fig. \ref{IQ}). Although the FoM could be different for another energy gate, the trend with the strip length is reproduced.
For this study we use the 9 strips that were read on each side which gives a panel of strip lengths. No condition on the event multiplicity was set as well as no condition on the coincidence between N-side and P-side so that the full strip length is used. Given that our detector is of nTD type, the uniformity in resistivity guarantees the low position dependence of the signal shape for a given particle at a given energy. It will thus have limited impact on the figures of merit and improve the statistical error. 
There are three groups of strip lengths: 50, 80 and 100~mm. \added{The two first ones correspond to N-side strips whereas the two last ones correspond to P-side strips. For the 80 mm long strips, we have data for both the N- and the P-side strips.} The different lengths correspond to a capacitance of 8, 13 and 18~pF respectively. There is more than a factor of 2 between the shortest and the longest strip in the detector. Fig.~\ref{fom} shows the figures of merit as a function of the strip length obtained for the N-side and P-side strips. 
When four strips are neighbors, the averaged value is shown by an horizontal line. The FoM for the N-side are slightly greater than that of the P-side, which is probably related to the better energy resolution obtained on the N-side. However for the strips with the same length (around 70-80~mm) the FoM are compatible between both sides. The trend is almost flat for proton/deuteron separation and slightly better for longer strips for deuteron/triton separation. This latter behavior is probably due to statistical errors that dominate the measurement in this case as the number of events with tritons is quite limited and even more in the shortest strip where the FoM is lower. 
On the P-side, the FoM slightly decreases with the strip length (from 2.2$\pm$0.1 to 1.9$\pm$0.15 on average for proton/deuteron separation) for both proton/deuteron and deuteron/triton separation.
For all cases, the FoM are above 1.5 so that the particle identification is achieved. In turn, the various lengths of the strips in the trapezoidal detector do not prevent from particle identification based on the pulse shape analysis technique.

\section{Conclusion and Perspectives}
We have described the first beam test results for the GRIT prototype detector with the first version of its ASIC preamplifier iPACI, a 9{\nbd}channel, 130~MHz - 12~keV FWHM \charge{} and \current{} output preamplifier.
The main goal of the study was to investigate the pulse shape analysis performance of the set-up. The focus was put on the effect of the different strip lengths from the trapezoidal detector. Although the strip capacitance varies by a factor of 2 along the detector, no strong effect was observed on the particle identification. A new version of the iPACI preamplifier improving the dynamic range and including more advanced blocks is being tested at the time of writing. This study opens the door for the next step using the full electronics chain to read the GRIT detector. 



\bibliographystyle{elsarticle-num} 
\bibliography{bibiPACI}

\end{document}